\documentclass[12pt,fleqn]{article}

\usepackage{epsfig}

\begin{document}

\title{\bf Pairwise entanglement in the XX model with a magnetic impurity}
\author{Hongchen Fu$^1$\thanks{h.fu@open.ac.uk} \
             Allan I Solomon$^1$\thanks{a.i.solomon@open.ac.uk} \
             Xiaoguang Wang$^2$\thanks{xgwang@isiosf.isi.it}
             \\
{\normalsize \it $^1$Quantum Processes Group, The Open University,} \\
{\normalsize \it Milton Keynes, MK7 6AA, United Kingdom}\\
{\normalsize \it $^2$Quantum Information Group, 
Institute for Scientific Interchange (ISI) Foundation,} \\
{\normalsize \it Viale Settimio Severo 65, I-10133 Torino, Italy }}
\maketitle

\begin{abstract}
For a 3-qubit Heisenberg model in a uniform magnetic field, the
pairwise thermal entanglement of any two sites is identical
due to the {\em exchange} symmetry of sites.  In this paper we 
consider the effect of a non-uniform magnetic field on the Heisenberg 
model, modeling a magnetic impurity on one site. Since pairwise 
entanglement is calculated by tracing out one of the three sites, 
the entanglement clearly depends on which site the impurity is located. 
When the impurity is located on the site which is traced out, that is, 
when it acts as an {\em external field} of the pair, the
 entanglement can be enhanced to the maximal value 1; while 
 when the field acts on a site of the pair the corresponding concurrence can only 
 be increased from 1/3 to 2/3.
\end{abstract}

\section{Introduction}

There is currently  an ongoing effort to study 
entanglement in multipartite systems, since such entangled states may 
provide a  valuable resource in quantum information
processing  \cite{Bennett}. Recently  entanglement in quantum operations
 \cite {EO0,EO1,EO2} and  in indistinguishable fermionic and bosonic
systems \cite{EI1,EI2,EI3} have been considered. 
Entanglement in   two-qubit states
has been  well studied in the literature. Various kinds of
three-qubit entangled states have also been  studied 
\cite{Dur,threeq,Rajagopal}, which have been shown to
possess advantages over two-qubit states in quantum teleportation \cite
{tele}, dense coding \cite{dense} and quantum cloning \cite{clone}. 

One interesting and natural type of entanglement, thermal
entanglement, was introduced and studied in the context of  the Heisenberg 
$XXX$ \cite{Arnesen},
$XX$ \cite{Wang1}, and $XXZ$ \cite{Wang2} models as well as the Ising 
model in a magnetic field \cite{Ising}. The Heisenberg interaction has 
been used to simulate a  quantum computer \cite{Loss}, and  can also
be realized in quantum dots \cite{Loss}, nuclear spins \cite{Kane}, 
electronic spins \cite{Vrijen} and optical lattices \cite{Moelmer}. By
suitable coding, the Heisenberg interaction can be used for quantum
computation \cite{Loss2}. Entanglement in
the ground state of the Heisenberg model has been discussed previously \cite{Oconnor}.
In an earlier note \cite{wfs} we presented 
an analytical study of pairwise entanglement in the  3-qubit Heisenberg
model in a {\em uniform} magnetic field and found that the
magnetic field can greatly enhance pairwise entanglement. 
Due to {\it exchange} symmetry in this cyclic model the entanglement of any two
sites is identical.  

In this paper we  consider the effect of
a magnetic impurity  on entanglement in the Heisenberg model. 
We find unsurprisingly that the effect of such an inhomogeneous 
magnetic field on the entanglement depends on which site the 
impurity is located, although in a non-intuitive way. When the
field may be considered as an {\em external field} of the pair, 
that is, when it is located on the site which is traced over, then 
it can enhance the entanglement to its  maximal
value, as measured by the concurrence.  When the field acts on a site of the pair the
concurrence can be increased from 1/3 to 2/3, but not to its maximal value 1.

\section{XX Heisenberg model with magnetic impurity}

We consider the 3-qubit XX Heisenberg model in a magnetic field
acting on the third site only. The Hamiltonian is \cite{Lieb}
\begin{equation}
  H=\frac{J}{2}\sum_{i=1}^3 \left(\sigma_{i}^x \sigma_{i+1}^x 
  + \sigma_{i}^y \sigma_{i+1}^y \right)
  +BJ \sigma_3^z, \label{hamiltonian}
\end{equation}
where we use $BJ$ rather than $B$ to denote the magnetic field.
The Hamiltonian (\ref{hamiltonian}) has eight distinct eigenvalues when
$B \neq 0$ 
\begin{eqnarray}
&& E_0 = -JB, \qquad
      E_1=\frac{J}{2}\left( 1+B_-\right), \nonumber\\
&& E_2=-J(1+B), \qquad
      E_3=-J(1-B), \nonumber\\
&& E_4=\frac{J}{2}\left( 1+B_+\right), \qquad
      E_5=\frac{J}{2}\left( 1-B_-\right), \nonumber\\
&& E_6=\frac{J}{2}\left( 1-B_+\right), \qquad
      E_7=JB,
\end{eqnarray}
where $B_{\pm} \equiv  (4B^2 \pm 4B + 9)^{1/2}$.
When $B=0$, the energy levels are degenerate
\begin{equation}
E_1 = E_7 =0, \quad E_1 = E_3 = E_4 =2J, \quad 
E_2 = E_5 =E_6 = -J.
\end{equation} 

In the  antiferromagnetic case ($J>0$), the ground state
is $E_2$, while in the ferromagnetic case ($J<0$),
the ground state is $E_4$. 

The corresponding non-degenerate, orthogonal 
eigenstates are
\begin{eqnarray}
&& |\phi_0\rangle = |000\rangle, \nonumber \\
&& |\phi_1\rangle = {\cal N}_1 \left(|100\rangle+|010\rangle
+a_1 |001\rangle \right), \nonumber\\
&& |\phi_2\rangle=2^{-1/2}\left(|010\rangle-|100\rangle\right), \nonumber \\
&& |\phi_3\rangle=2^{-1/2}\left(|101\rangle-|011\rangle\right), \nonumber\\
&& |\phi_4\rangle = {\cal N}_4 \left(a_4|110\rangle+|101\rangle
+|011\rangle \right), \nonumber\\
&& |\phi_5\rangle = {\cal N}_5 \left(|100\rangle+|010\rangle
+a_5 |001\rangle \right), \nonumber\\
&& |\phi_6\rangle = {\cal N}_6 \left(a_6|110\rangle+|101\rangle
+|011\rangle \right), \nonumber \\
&& |\phi_7\rangle = |111\rangle,
\end{eqnarray}
where
\begin{eqnarray}
&& a_1=-\frac{1}{2}+\frac{1}{2} B_- + B, \qquad
      a_5=-\frac{1}{2}-\frac{1}{2} B_- +B, \nonumber\\
&& a_4=-\frac{1}{2}+\frac{1}{2} B_+ -B, \qquad
      a_6=-\frac{1}{2}-\frac{1}{2} B_+ -B, 
\end{eqnarray}
and ${\cal N}_i = (2+a_i^2)^{-1/2}$ ($i=1,4,5,6$) are normalization constants.

It is interesting to note that the eigenvalues transform  under
 $B\leftrightarrow -B$ by
\begin{equation} \label{eigensym}
E_0\leftrightarrow E_7, \quad 
E_1\leftrightarrow E_4, \quad
E_2\leftrightarrow E_3, \quad
E_5\leftrightarrow E_6,
\end{equation}
and so the $a_i$'s transform
by
$a_1\leftrightarrow a_4, \quad a_5\leftrightarrow a_6.$
This leads to invariance of the  entanglement under $B\leftrightarrow -B$.

The density operator $\rho(T)$ at temperature $T$ can be written as
\begin{equation}
\rho(T)=\frac{1}{Z} \sum_{i=0}^7 e^{-\beta E_i}
|\phi_i\rangle \langle \phi_i|,
\end{equation}
where $\beta=1/kT$ and $Z$ is the partition function
\begin{eqnarray}
Z&=&\mbox{tr}\left(e^{-\beta H}\right)= \sum_{i=0}^7 e^{-\beta E_i} \\
 & = &  2 (1+ e^{J\beta}) \cosh(J\beta B)+
 2e^{-J\beta/2} \left[
 \cosh\left(\frac{1}{2}J\beta B_+ \right)+ 
 \cosh\left(\frac{1}{2}J\beta B_-\right)\right].
\end{eqnarray}

\section{Concurrence of pairwise entanglement}

The easiest way to calculate the entanglement is by means of the 
{\em concurrence} ${\cal C}$ \cite{conc} between a pair of qubits, 
which is defined as
\begin{equation}
{\cal C} = \max \left\{ \lambda _1-\lambda _2-\lambda _3-\lambda_4,0\right\},  
\label{eq:c1}
\end{equation}
where the quantities $\lambda _i$ are the square roots of the eigenvalues of
the operator 
\begin{equation}
\varrho=\rho (\sigma _1^y\otimes \sigma _2^y)\rho^{*}(\sigma
_1^y\otimes \sigma_2^y)  \label{eq:c2}
\end{equation}
in descending order; $\rho$ is the density operator 
of the pair and it can be
either pure or mixed.
The entanglement of formation is a monotonic function of the concurrence 
${\cal C}$, varying between a minimum of zero for ${\cal C}=0$, 
and a maximum of 1 for ${\cal C}=1$. 

We now derive the concurrence for  any pair of sites in our model. 
Due to  symmetry under the exchange of  sites 1 and 2, the 
entanglement between sites 1 and 3 is  the same 
as that between  sites 2 and 3, and so  we need only consider 
 entanglement between sites 1 and 3, and between sites 
1 and 2. 

Taking the trace over the second (third) site, we can 
obtain the reduced density operator $\rho_{13}$($\rho_{12}$) 
of the sites 1 and 3 (1 and 2).  Both $\rho_{12}$
and $\rho_{13}$ take the following form
\begin{equation}
\rho=\frac{1}{Z}\left(
\begin{array}{cccc}
 u & & & \\
   & w_1 & y & \\
   & y & w_2 & \\
   &   &   & v 
\end{array}\right)
\end{equation}
Here, for $\rho_{12}$, the nonzero matrix elements are given by 
\begin{eqnarray}
y &=& {\cal N}_1^2 e^{-\beta E_1}+{\cal N}_4^2 e^{-\beta E_4}
   +{\cal N}_5^2 e^{-\beta E_5} 
   + {\cal N}_6^2 e^{-\beta E_6} \nonumber\\
   && -\frac{1}{2} e^{-\beta E_2} -\frac{1}{2} e^{-\beta E_3} \nonumber\\
w_1=w_2 &=& {\cal N}_1^2 e^{-\beta E_1}+{\cal N}_4^2 e^{-\beta E_4}
   +{\cal N}_5^2 e^{-\beta E_5} 
   + {\cal N}_6^2 e^{-\beta E_6} \nonumber \\
   && +\frac{1}{2} e^{-\beta E_2} +\frac{1}{2} e^{-\beta E_3}\nonumber \\
u &=& e^{-\beta E_0}+a_1^2 {\cal N}_1^2 e^{-\beta E_1}+
   a_5^2 {\cal N}_5^2 e^{-\beta E_5}, \nonumber \\
v &=& e^{-\beta E_7}+a_4^2{\cal N}_4^2 e^{-\beta E_4}  +
  a_6^2 {\cal N}_6^2 e^{-\beta E_6}.
\end{eqnarray}
while for the $\rho_{13}$ case, we have
\begin{eqnarray}
y &=& a_1{\cal N}_1^2 e^{-\beta E_1}+a_4 {\cal N}_4^2 e^{-\beta E_4}
   +a_5 {\cal N}_5^2 e^{-\beta E_5} 
   + a_6 {\cal N}_6^2 e^{-\beta E_6}, \nonumber\\
w_1 &=& a_1^2 {\cal N}_1^2 e^{-\beta E_1}+\frac{1}{2} e^{-\beta E_3}
   +{\cal N}_4^2 e^{-\beta E_4}
   +a_5^2 {\cal N}_5^2 e^{-\beta E_5} 
   + {\cal N}_6^2 e^{-\beta E_6} \nonumber\\
w_2 &=&  {\cal N}_1^2 e^{-\beta E_1}+\frac{1}{2} e^{-\beta E_2}
   + a_4^2 {\cal N}_4^2 e^{-\beta E_4}
   +{\cal N}_5^2 e^{-\beta E_5} 
   + a_6^2 {\cal N}_6^2 e^{-\beta E_6} \nonumber\\   
u &=& e^{-\beta E_0}+ {\cal N}_1^2 e^{-\beta E_1}+\frac{1}{2}e^{-\beta E_2}+
   {\cal N}_5^2 e^{-\beta E_5}, \nonumber\\
v &=& e^{-\beta E_7}+\frac{1}{2}e^{-\beta E_3}+
  +{\cal N}_4^2 e^{-\beta E_4}  +
  {\cal N}_6^2 e^{-\beta E_6}.
\end{eqnarray}

The concurrence has the form
\begin{equation}
{\cal C}=\frac{2}{Z} \max \left\{ |y|-\sqrt{uv}, \ 0 \right\}.
\end{equation}
The system is entangled when ${\cal C}>0$, and maximally entangled when
${\cal C}=1$. The exchange interaction constant $J$ and the temperature 
$T$ always appear in the form $J/kT$ in the concurrence and thus we 
can define the {\em scaled temperature}
$\tau \equiv kT/|J| \geq 0$. 
The concurrence is a function of $\tau$ and $B$. 

From Eqs.\,(\ref{eigensym}) it is easy to see that 
$y\to y$ and $u\leftrightarrow v$ when $B\to -B$. This means that
the concurrence is invariant under $B\leftrightarrow -B$;
\begin{equation}
{\cal C}(\tau, B)={\cal C}(\tau, -B).
\end{equation}
We therefore only consider the case $B\geq 0$ case hereafter.

\section{Discussion and results}


\subsection{${\cal C}_{12}$}

We first consider  the entanglement between  sites 1 and 2.  
In Fig.\,1 and 2  we give plots of the concurrence of $\rho_{12}$ 
against $\tau$ and $B$.
We know that  entanglement appears only in the 
antiferromagnetic case ($0<\tau\leq 1.27$) when $B=0$  
\cite{wfs} (also see Figure 1). From Fig. 1 and 2 we see that,
when the magnetic impurity is located on the third site, both the 
antiferromagnetic and ferromagnetic cases are entangled
in the range $0 < \tau \leq \tau_0$, where $\tau_0$ depends on $B$. 

Fig.\,1 also suggests that the concurrence ${\cal C}_{12}$ goes 
to $1$, namely, the sites 1 and 2 reach  maximal entanglement,
when $\tau \to 0$ for large enough $B$, in both the antiferromagnetic
and ferromagnetic cases. This 
fact can be shown analytically as follows. 

Consider first the antiferromagnetic case $(J>0)$. In this case 
$E_2$ is the ground state; that is, $E_2 - E_i <0 $ for all 
$i\neq 2$ and thus
$e^{-\beta E_2} \gg e^{-\beta E_i}$ for
$i \neq 2$ in the limit $\tau \to 0$. 
Note that all ${\cal N}_i$ and $a_i$ are finite. Then we have 
\begin{equation}
y\to \frac{1}{2} e^{-\beta E_2}, \qquad
Z\to e^{-\beta E_2} ,\qquad 
\frac{u}{Z} \to 0, \qquad
\frac{v}{Z} \to 0,
\end{equation}
namely, ${\cal C}_{12} \to 1$ when $\tau \to 0$. 

For the  ferromagnetic case ($J<0$), 
one can check that $E_4-E_i < 0$ for all $i \neq 4$ and 
$e^{-\beta E_4} \gg e^{-\beta E_i}$ ($i\neq 4$)
in the limit $\tau\to 0$.
Then we have 
\begin{equation}
y\to {\cal N}_4^2  e^{-\beta E_4}, \qquad
Z\to e^{-\beta E_4} ,\qquad 
\frac{u}{Z} \to 0, \qquad
\frac{v}{Z} \to a_4^2 {\cal N}_4^2,
\end{equation}
namely, 
\begin{equation}
    {\cal C}_{12} \to 2{\cal N}_4^2 
    =\frac{2}{2+ a_4^2}. 
\end{equation}
when $\tau \to 0$.
In the limit $B\to \infty$, $a_4\to 0$ and therefore
${\cal C}_{12} \to 1$.
In the limit $B\to +0$, but $B\gg \tau$, 
 ${\cal C}_{12} \to 2/3$. 
 
It is interesting to note that, when $B=0$, 
 ${\cal C}_{12} \to 1/3$ in the limit $\tau \to 0$ \cite{wfs}. In this case
 the ground state is 3-fold degenerate and the approximation we
 used above is not valid. This again indicates the role of degeneracy
 in entanglemant. 

\subsection{${\cal C}_{13}$}

We consider the entanglement between sites 1 and 3. 
>From Fig. 3 and 4 we see that: 

1. In contrast to the 1-2 case, the concurrence increases to a
maximum with  increasing $B$ and then decreases.  
The lower $\tau$, the smaller $B$ at which the concurrence 
reaches its maximum value.

2. For small $B$,  entanglement occurs only in the  ferromagnetic
case ($J<0$), while for large enough $B$  (e.g. $B=10$), 
entanglement occurs in both the antiferromagnetic
and  ferromagnetic cases, but it is very weak. 

Fig.\,4 suggests that the maximal entanglement occurs in the
ferromagnetic case when
$\tau \to 0$ and $B$ is also much smaller than 1. In this case,
$E_1$ is very close to the ground state $E_4$ and 
$\exp(-\beta E_4)$ and $\exp(-\beta E_1)$  are much bigger
than others. We can also check that
\begin{equation}
\frac{e^{-\beta E_4}}{e^{-\beta E_1}} \sim
\exp\left(\frac{2}{3}\frac{B}{\tau}\right) \geq 1,
\end{equation}
and that
\begin{equation}
{\cal N}_1 \sim {\cal N}_4 \sim \frac{1}{3}, \qquad 
a_1 \sim a_4 \sim 1.
\end{equation}
The concurrence is then given approximately by
\begin{equation}
C_{13} \sim \frac{2}{3}\left[1-
\frac{\exp\left(\frac{1}{3}\frac{B}{\tau}\right)}{
1+\exp\left(\frac{2}{3}\frac{B}{\tau}\right)}
\right],
\end{equation}
from which we conclude that the maximal concurrence is 
$2/3$ when $B$ is much bigger than $\tau$ and much smaller
than 1.  

\vspace{0.3cm}
In summary, we list our results in the following table.

\begin{tabular}{|c|l|l|} \hline
        & Maximal concurrence & Entanglement ranges \\ \hline \hline
B=0 & 1/3 & Antiferromagnetic case only \\ \hline
12   & 1, when $|\tau|\to 0$ and  & In both ferromagnetic and antiferromagnetic
\\ & $B $ is big enough. & cases  \\ \hline
13    & 2/3  & When $B$ is small, only the ferromagnetic case is \\
& for antiferromagnetic    & entangled. When $B$ is big enough, both\\
& case and $\tau \ll B \ll 1$ &  the antiferromagnetic and ferromagnetic cases\\
& &  are entangled, but the entanglement is very weak.\\ \hline
\end{tabular}

\section{Conclusion}

In this paper we 
considered the effect of a non-uniform magnetic field on the Heisenberg 
XX model, modeling a magnetic impurity on only one site. 
In contrast to the uniform magnetic field case \cite{wfs} where the
pairwise thermal entanglement of any two sites is identical
due to the {\em exchange} symmetry of sites,
the entanglement due to a  non-uniform magnetic field clearly depends 
on which site the impurity is located. 
When the impurity is located on the site which is traced out, that is, 
when it acts as an {\em external field} of the pair, the concurrence corresponding to the 
entanglement can be enhanced to the maximal value 1 from 1/3; while 
when the field acts on a site of the pair the concurrence can only 
be increased from 1/3 to 2/3. Maximal entanglement is
achieved when the temperature tends to zero. 

In \cite{wfs}, the entanglement was related  to 
the degeneracy of the system. In the present model, the magnetic 
field removes all the degeneracy of the energy levels present when $B=0$
and the entanglement is thus greatly enhanced.

\section{Acknowledgement}

X. Wang is supported by European project Q-ACTA.


\newpage

\begin{figure}
\begin{center}
\epsfxsize=12cm
\epsffile{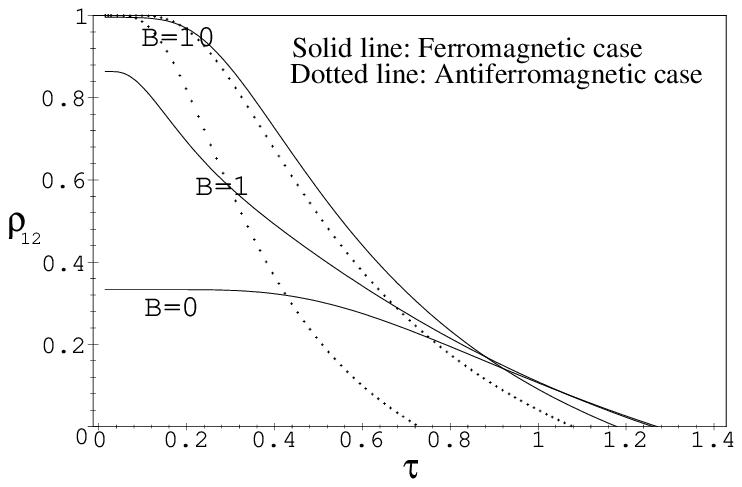} 
\end{center}
\caption{Concurrence ${\cal C}_{12}$ against $\tau$ 
for different magnetic field $B=0,1,10$.}
\end{figure}

\begin{figure}
\begin{center}
\epsfxsize=12cm
\epsffile{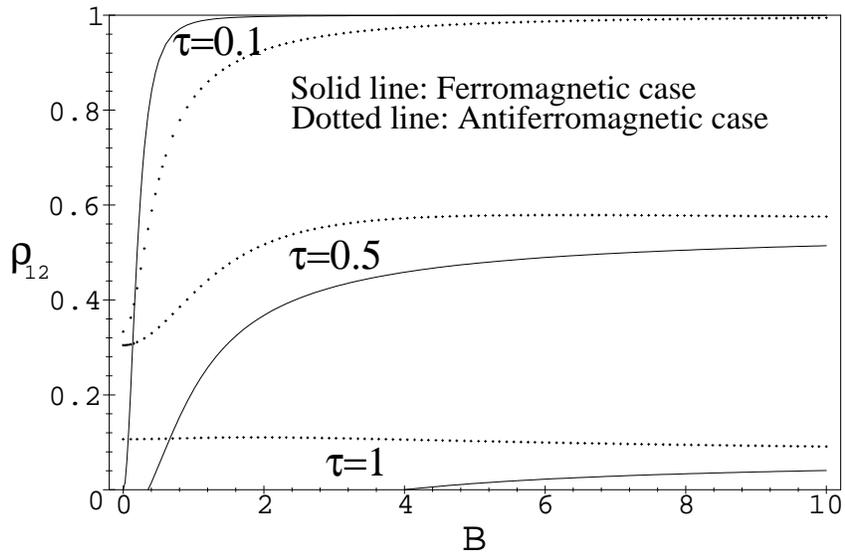} 
\end{center}
\caption{Concurrence ${\cal C}_{12}$ against $B$ 
at different temperature $\tau=0.1,0.5$ and $1$.}
\end{figure}


\begin{figure}
\begin{center}
\epsfxsize=12cm
\epsffile{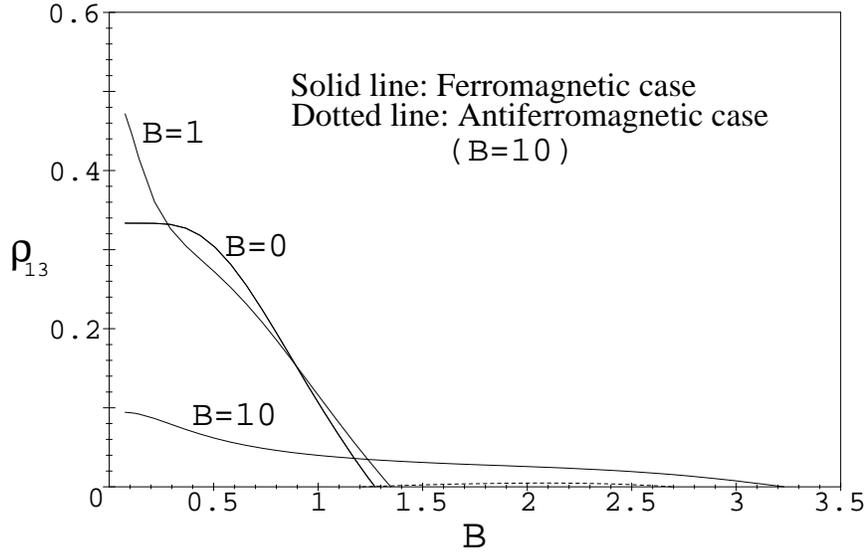} 
\end{center}
\caption{Concurrence ${\cal C}_{13}$ against $\tau$ 
for different magnetic field $B=0,1,10$.
For antiferromagnetic case (dotted line) with $B = 10$,
the entanglement occurs although it is very week.}
\end{figure}

\begin{figure}
\begin{center}
\epsfxsize=12cm
\epsffile{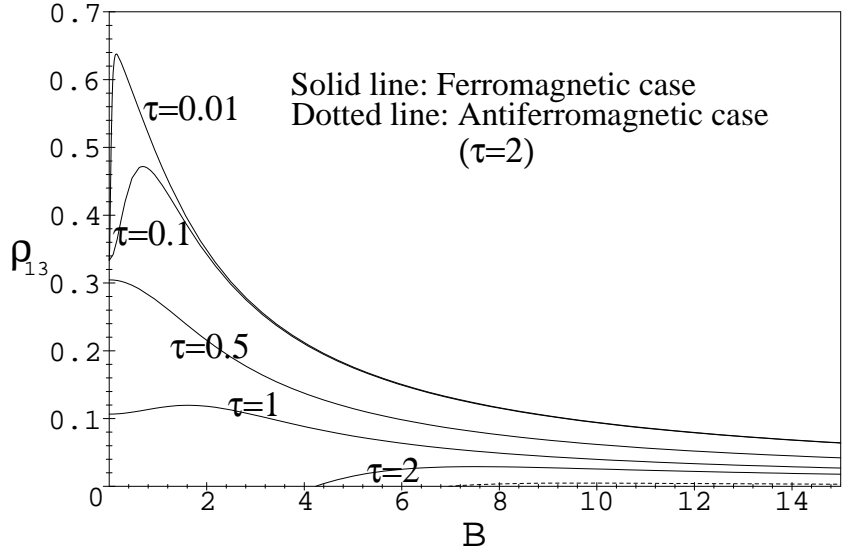}
\end{center}
\caption{Concurrence ${\cal C}_{13}$ against $B$ 
at different temperature. For antiferromagnetic case (dotted line),
$\tau = 2$.}
\end{figure}

\end{document}